# Sizable suppression of magnon Hall effect by magnon damping in $Cr_2Ge_2Te_6$

Ysun Choi[1,2,*], Heejun Yang[1,2,*], Jaena Park[1,2], and Je-Geun Park[1,2,$]

[1]Center for Quantum Materials, Seoul National University, Seoul 08826, Republic of Korea

[2]Department of Physics and Astronomy, Seoul National University, Seoul 08826, Republic of Korea

* These authors made equal contributions.

[$]jgpark10@snu.ac.kr

## Abstract

Two-dimensional (2D) Heisenberg honeycomb ferromagnets are expected to have interesting topological magnon effects as their magnon dispersion can have Dirac points. The Dirac points are gapped with finite second nearest neighbor Dzyaloshinskii-Moriya interaction, providing nontrivial Berry curvature with finite magnon Hall effect. Yet, it is unknown how the topological properties are affected by magnon damping. We report the thermal Hall effect in $Cr_2Ge_2Te_6$, an insulating 2D honeycomb ferromagnet with a large Dirac magnon gap and significant magnon damping. Interestingly, the thermal Hall conductivity in $Cr_2Ge_2Te_6$ shows the coexisting phonon and magnon contributions. Using an empirical two-component model, we successfully estimate the magnon contribution separate from the phonon part, revealing that the magnon Hall conductivity was 20 times smaller than the theoretical calculation. Finally, we suggest that such considerable suppression in the magnon Hall conductivity is due to the magnon damping effect in $Cr_2Ge_2Te_6$.

## I. INTRODUCTION

Topological physics is arguably the most fundamental and profound discovery made in condensed matter physics over the past few decades. For example, the quantum geometrical factor, now known as Berry curvature, has been considered essential for describing novel quantum phenomena such as the Aharonov-Bohm effect, quantum Hall effect, anomalous Hall effect, etc. [1–5]. Recently, theoretical extensions were made for bosonic quasiparticles, especially magnon [6], an elementary excitation of a spin system. Like the fermionic counterpart, the Berry curvature of magnons can also induce transverse velocity on the magnon wave packet, resulting in the magnon Hall effect [7]. Kagome ferromagnet was a promising candidate for the magnon Hall effect [6] since the nearest neighbor (NN) bonds allow Dzyaloshinskii-Moriya (DM) interaction [8,9]. When this DM vector is parallel to the magnetization, a gap opens up in the magnon band crossing points, producing nontrivial Berry curvature [10]. The first experimental report of the magnon Hall effect was made in $Lu_2V_2O_7$, an insulating pyrochlore ferromagnet [11]. Since then, several new examples of the magnon Hall effect have been found in other kagome magnets [12–18].

Honeycomb lattice has recently attracted more attention as a new system of hosting the topological magnon since the Heisenberg Hamiltonian produces Dirac-like linear band crossing points of magnon [19], just like the electronic band structure of graphene. As shown in Fig. 1(a), the honeycomb lattice also allows DM interaction for the second-NN bonds [8,9], which can open a gap at Dirac crossing points, resulting in nontrivial Berry curvature [20,21]: which is a precise analogy to the Haldane model for graphene [22]. The sizable magnon Hall effect was recently experimentally identified in $VI_3$, an insulating honeycomb ferromagnet with a DM interaction of 0.2 meV [23].

But there is a clear distinction to be made between the magnon topology and the electronic counterpart. Unlike the electronic bands that are typically coherent Bloch states, magnon damping due to multi-particle interaction is inevitable for many spin systems [24–30]. Thus, understanding this magnon damping is a critical problem for the further development of magnon topology. Unfortunately, the current formula of the magnon Hall conductivity is based on linear spin-wave theory (LSWT) [7], with little regard for the consequence of higher-order terms. Therefore, it is crucial to investigate how magnon damping affects magnon transport [31–33], and we introduce $Cr_2Ge_2Te_6$ (CGT) as an ideal candidate.

CGT is a two-dimensional insulating van der Waals (vdW) magnet consisting of Cr honeycomb layers, exhibiting extremely soft ferromagnetic behavior in bulk with a nearly absent coercive field in magnetization [34]. Several measurements, including Raman scattering, thermal expansion, and thermal conductivity, reported significant spin-phonon interaction in CGT [35–38], essential to spintronics applications. Its magnetic Hamiltonian ($H_m$) was given by recent neutron studies as follows [39,40],

$$H_m = \sum_{\langle ij \rangle_n, l} J_n \mathbf{S}_{i,l} \cdot \mathbf{S}_{j,l} + \sum_{\langle ij \rangle_n, \langle lm \rangle} J_{c,n} \mathbf{S}_{i,l} \cdot \mathbf{S}_{j,m}$$
$$+ \sum_{\langle ij \rangle_2, l} \mathbf{D} \cdot (\mathbf{S}_{i,l} \times \mathbf{S}_{j,l}) - K \sum_{i,l} (S_{i,l}^z)^2 - g\mu_B \mu_0 H \sum_{i,l} S_{i,l}^z, \quad (1)$$

where $\mathbf{D}$ is the DM vector [41] and the parameters are summarized in Table I. As shown in Fig. 1(b), we can note that the calculated magnon Hall conductivity ($\kappa_{xy}^{calc}$) based on $H_m$ has a single peak around Curie Temperature ($T_C$), and the size of $\kappa_{xy}^{calc}$ is proportional to $|\mathbf{D}|$. In the case of $VI_3$, where the form of $H_m$ is still valid, the observed magnon Hall conductivity was effectively explained by this $\kappa_{xy}^{calc}$ following a parallel manner [23]. Given that $|\mathbf{D}|$ is comparable between CGT and $VI_3$, we can expect that the magnon Hall effect in CGT should be similar to VI$_3$. But, contrary to VI$_3$, CGT hosts strong magnon damping, another point to be considered for a proper understanding of the magnon Hall effect. The latest neutron study [40] claimed that the significant exchange-striction type spin-phonon coupling ($H_{mp}$) on $J_1$ should be considered for CGT, where $\bm{r}_{ij}$ denotes displacement of Cr ions between site $i$ and $j$,

$$H_{mp} = \sum_{\langle ij \rangle, l} (\bm{S}_{i,l} \cdot \bm{S}_{j,l}) \left( \frac{\partial J_1}{\partial r_{ij}} \cdot \bm{r}_{ij} \right). \tag{2}$$

Therefore, CGT hosts both large DM interaction and magnon damping simultaneously and hence can be an appropriate example to study how magnon damping affects the magnon Hall effect.

In this paper, we report the experimental measurement of the thermal Hall effect in CGT. The temperature dependence of thermal Hall conductivity ($\kappa_{xy}$) exhibits multiple peaks: a sizeable positive peak near 20 K and smaller peaks around $T_C$ with a sign change. The magnetic field dependence of $\kappa_{xy}(H)$ follows the magnetization ($M(H)$) at a low-temperature region, and we observed an additional negative component in $\kappa_{xy}(H)$ around $T_C$. We applied an empirical two-component model consisting of a positive magnetization-like term ($\alpha M(H)$) and negative field-linear term ($-\beta H$), i.e., $\kappa_{xy}(H) = \alpha M(H) - \beta H$ ($\alpha, \beta \geq 0$). We found that this model fits our $\kappa_{xy}(H)$ extremely well for the overall temperature range, and we assigned the positive (negative) term as phonon (magnon) contribution. We found that the negative magnon term was 20 times smaller than $\kappa_{xy}^{calc}$ obtained from LSWT calculation, from which we suggest that the magnon damping effect suppresses the overall size of the magnon Hall effect.

## II. EXPERIMENTAL DETAILS AND RESULTS

Single crystals of CGT were synthesized using a self-flux method, as described in Ref. [42]. The out-of-plane magnetization $M(T)$ shows typical ferromagnetic behavior with $T_C$ at 67 K as determined from a sharp peak in $dM/dT$ (Fig. 1(c)), consistent with the previous reports [38,42]. The thermal Hall measurement was performed by the conventional steady-state method under the magnetic field parallel to the out-of-plane direction of the CGT sample. As shown in Fig. 1(d), a heater attached to a plate-like CGT sample generates heat current along the *x* direction, while the other three thermometers ($T_1$, $T_2$, and $T_3$) measure temperature differences for each *x* and *y* directions ($\Delta T_x$ and $\Delta T_y$). We employed SrTiO$_3$ capacitive thermometers [43] to minimize the calibration error due to the high magnetic field: the dielectric constant of SrTiO$_3$ shows almost negligible field effect [44]. We also antisymmetrized $\Delta T_y$ with opposite magnetic field directions using the following relation $\Delta T_y^{\mathrm{asym}} = \frac{\Delta T_y(+H) - \Delta T_y(-H)}{2}$. This is a common procedure for extracting small intrinsic Hall signals ($\Delta T_y^{\mathrm{asym}}$) in $\Delta T_y$ from larger artifacts that might arise from misalignment between two transverse contacts ($T_2$ and $T_3$) [15,45–47]. The finally obtained $\Delta T_x$ and $\Delta T_y^{\mathrm{asym}}$ are then converted into longitudinal thermal conductivity ($\kappa_{xx}$) and $\kappa_{xy}$ by the Fourier's law of thermal conduction, respectively.

As shown in Fig. 1(e), our $\kappa_{xx}(T)$ data in zero field reproduce several key features reported in previous reports; a single peak around 25 K, a downward cusp at $T_C$, and flat temperature dependence for $T > T_C$ [38,48]. Phonons are natural heat carriers in insulators like CGT, and a single peak around 20 K has been frequently seen in typical $\kappa_{xx}(T)$ data [49]. But the $\kappa_{xx}(T)$ data of CGT show one distinct feature different from most of the $\kappa_{xx}(T)$ data due to phonons alone. Typical phonon contributions show a smoothly decreasing curve at a high-temperature range, rather than a sharp cusp as seen in the $\kappa_{xx}(T)$ data of CGT. Recently, spin fluctuations were proposed as a possible explanation for such an abrupt cusp behavior in $\kappa_{xx}(T \geq T_C)$ since spin fluctuations can provide an additional scattering source of phonons via a spin-phonon coupling [38,48,50–53]. In other words, spin fluctuations can suppress $\kappa_{xx}$ further from its original behavior of the Debye-Callaway model [54], resulting in a cusp around the magnetic phase transition. Thus, the flat $\kappa_{xx}(T \geq T_C)$ in CGT implies that CGT hosts significant spin-phonon coupling, consistent with Raman [35] and thermal expansion [36,37] studies.

Fig. 2(a) presents the temperature dependence of $\kappa_{xy}(T)$, measured under the magnetic field of 1 T. A glance reveals that $\kappa_{xy}(T)$ shows a distinct positive peak near 20 K, which seems to converge toward zero rapidly as the temperature increases. However, upon a more careful examination of the data, we observed a small negative peak near $T_C$ as shown in the inset of Fig. 2(a). Then $\kappa_{xy}(T)$ changes its sign once again and becomes positive, ultimately converging to 0 for $T \gg T_C$.

Figs. 2(b) and (c) show the magneto-thermal conductivity ($\frac{\Delta\kappa_{xx}(H)}{\kappa_{xx}(0)}$) defined as $\frac{\kappa_{xx}(H)-\kappa_{xx}(0)}{\kappa_{xx}(0)}$, $M(H)$, and $\kappa_{xy}(H)$. The isothermal $M(H)$ exhibits soft ferromagnetic behavior with negligible hysteresis and a saturation field ($H_S$) around 0.2 T, as reported before [34]. At the same time, $\frac{\Delta\kappa_{xx}(H)}{\kappa_{xx}(0)}$ shows monotonic increasing behavior at $H \geq H_S$ for the overall temperature range, which was commonly observed in other Cr based honeycomb vdW magnets in a ferromagnetically ordered phase [53,55]. Considering the dominant phonon contribution in $\kappa_{xx}$, this increasing $\frac{\Delta\kappa_{xx}(H)}{\kappa_{xx}(0)}$ can be explained by a reduced phonon scattering rate due to suppressed spin fluctuations (or magnon population) [53]. Additionally, we noted nearly flat but slowly increasing behavior in $\frac{\Delta\kappa_{xx}(H)}{\kappa_{xx}(0)}$ with $H < H_S$ for $T < T_C$. We suppose that this kind of feature can arise from magnetic domain walls [56], which can scatter phonons and thus reduce a phonon mean free path [57–59] for $H < H_S$. On the other hand, we also found that $\kappa_{xy}(H)$ behaves quite similarly to $M(H)$ for $T \ll T_C$. Interestingly, as shown in Fig. 2(c), $\kappa_{xy}(H)$ starts to deviate from $M(H)$ for $T \sim T_C$: it seems to acquire a negative linear component accompanying the sign change. This negative term in $\kappa_{xy}(H)$ gets diminished as the temperature increases, and $\kappa_{xy}(H)$ becomes eventually similar to $M(H)$ again.

We do not think that a single heat carrier model, either phonon or magnon, can explain the above distinctive features in $\kappa_{xy}$: i.e., the multiple peaks in $\kappa_{xy}(T)$ and the coexistence of positive and negative terms. Instead, we suggest that it is natural to consider at least two types of transverse heat carriers for CGT. Following this idea, we first need to decompose the measured $\kappa_{xy}(T)$ into contributions due to each transverse heat carrier. For this, we used the following empirical formula $\kappa_{xy}(H) = \alpha M(H) - \beta H$ ($\alpha, \beta \geq 0$), where both $\alpha$ and $\beta$ are assumed to be fitting parameters. It then consists of a positive magnetization-like term $\alpha M(H)$ and a negative field-linear term $-\beta H$. Surprisingly, this simple empirical two-component model shows excellent agreement with the experimental data of $\kappa_{xy}(H)$ for the overall temperature range [60], as shown in solid black curves in the lowest panels of Figs. 2(b) and (c).

From the above fitting result, we can obtain the temperature dependence of both $\alpha M(H)$ and $-\beta H$ terms under the magnetic field of 1 T (Fig. 3(a)). We observe that the $\alpha M(H)$ term displays a prominent positive peak at 20 K, with another much smaller peak around 70 K. On the other hand, the extracted $-\beta H$ term has finite values only around $T_C$ with a single negative peak of 1 mWK$^{-1}$m$^{-1}$ (inset of Fig. 3(a)).

## III. DISCUSSION

Phonon ($\kappa_{xy}^{ph}$) and magnon ($\kappa_{xy}^{mag}$) are natural candidates for the thermal Hall effect in CGT. In addition, recent theoretical studies showed that magnon-phonon hybridized excitation ($\kappa_{xy}^{mag-ph}$) could also contribute to thermal Hall conductivity [61–64], which can be characterized by finite gap opening on crossing points between magnon and phonon bands [26]. However, such gap opening was not reported in the recent neutron spectrum of CGT [40], which implies $\kappa_{xy}^{mag-ph}$ contribution could be negligible compared to $\kappa_{xy}^{ph}$ and $\kappa_{xy}^{mag}$. Therefore, we assumed $\kappa_{xy}$ as a sum of the two contributions, i.e., $\kappa_{xy} = \kappa_{xy}^{ph} + \kappa_{xy}^{mag}$ [18,23]. According to our successful decomposition of $\kappa_{xy}$ to $\alpha M(H)$ and $-\beta H$ terms, we can safely assume that each term represents contributions due to phonons or magnons, respectively. Interestingly, recent extensive studies on cuprates reported some common properties for the phonon Hall effect. First, the temperature dependence of $\kappa_{xy}^{ph}(T)$ is very similar to $\kappa_{xx}(T)$ [65–68]. Since $\kappa_{xx}(T)$ is generally considered to arise mainly from phonons, such similarity between $\kappa_{xy}^{ph}(T)$ and $\kappa_{xx}(T)$ implies that they should share the same origin, that is phonon. Second, the Hall angles ($|\kappa_{xy}/\kappa_{xx}|$) for such reports are found to be around $3 \times 10^{-4}$ for the phonon Hall effect [68]. Here, we would like to focus on the overall temperature behavior of $\alpha M(H)$ term as it shows a dominant peak around 20 K, similar to $\kappa_{xx}(T)$ of CGT. In addition, the magnitude of $|\kappa_{xy}/\kappa_{xx}|$ in CGT is about $2.2 \times 10^{-4}$ at the 20 K peak position, which is comparable to the typical values expected for the phonon case.

For the magnon Hall effect, $\kappa_{xy}^{mag}(T)$ is expected to show a peak close to $T_C$ as often seen in other ferromagnetic insulators [11,12,23]. Under the magnetic field, the size of $\kappa_{xy}^{mag}(H)$ is expected to get diminished with the increasing field in the low-temperature range since the magnon band energy will be shifted upwards, lowering the magnon population [11,12,23]. However, our $\kappa_{xy}(H)$ data at 15 K keep increasing gradually even in higher magnetic fields up to 5 T (Fig. S4), which is hard to be explained by magnons. Hence, we judge that the $\alpha M(H)$ term cannot be easily explained by a magnon scenario. On the other hand, previous experimental studies on ferromagnets reported that $\kappa_{xy}^{mag}(H)$ became almost linear in a magnetic field around $T_C$ [11,12]. This also supports that the $-\beta H$ term is more likely to originate from magnons. Therefore, we can possibly conclude that the positive $\alpha M(H)$ term represents the phonon Hall effect, $\kappa_{xy}^{ph} = \alpha M(H)$, whereas the remaining negative $-\beta H$ term is due to the magnon Hall effect, $\kappa_{xy}^{mag} = -\beta H$.

From now on, we want to present detailed discussions for each $\kappa_{xy}^{ph}$ and $\kappa_{xy}^{mag}$ terms obtained using the above two-component model. Several theoretical attempts for $\kappa_{xy}^{ph}$ were made by introducing various ideas: Berry curvature of phonon bands [69–71], skew-scattering from rare-earth ions [72], structural domains [73], and complex kinetic theories [74–76]. Unfortunately, it is hard to determine the exact mechanism for $\kappa_{xy}^{ph}$ in CGT at the moment. Instead, we can speculate that the phonon Hall effect in CGT is due to the secondary effect of spin-phonon coupling [45,77]. On the other hand, we notice that our $\kappa_{xy}^{ph}$ shows a small extra hump around 70 K, which is counter-intuitive to the smooth exponentially decaying behavior as predicted in the recent phenomenological theory [78]. We think that this anomaly comes from a significant magnetoelastic coupling combined with a ferromagnetic phase transition, as shown in recent thermal expansion studies [36,37].

We compare our $\kappa_{xy}^{mag}$ with theoretical magnon Hall conductivity ($\kappa_{xy}^{calc}$) as obtained using LSWT [60]. Interestingly, the inset of Fig. 3(a) shows that the temperature dependence of $\kappa_{xy}^{mag}(T)$ and $\kappa_{xy}^{calc}(T)$ are quite similar to each other. However, it is noteworthy that the size of $\kappa_{xy}^{mag}(T)$ is 20 times smaller than $\kappa_{xy}^{calc}(T)$. One possible explanation for this huge size difference between the experimental and theoretical results could be that somehow the DM interaction is grossly overestimated from the neutron analysis [39,40]. The latest study suggests that about half of the reported DM value is appropriate for the $Cr_2X_2Te_6$ (X=Si, Ge) family [79]. Noting such perspective, we also calculated $\kappa_{xy}^{calc}$ as a function of DM interaction at the temperature of 64 K, a peak position of $\kappa_{xy}^{calc}(T)$ (Fig. 3(b)). We can clearly see that an exceptionally tiny value of |**D**| = 0.005 meV, two orders of magnitude smaller than the reported DM value, can explain the experimental $\kappa_{xy}^{mag}$ data with a typical size of 1 mWK$^{-1}$m$^{-1}$. A big problem with this explanation is that our simulated magnon bands using |**D**| = 0.005 meV cannot give a visible band gap at the magnon band crossing point, which conflicts with the neutron studies displaying a clear gap of 1 meV order (Fig. 3(c)). Thus, we can reject the simple idea of smaller |**D**| value, requiring a new explanation.

Notably, the latest neutron study showed that the exchange-striction type spin-phonon coupling ($H_{mp}$) plays an essential role in explaining the significant magnon damping in CGT [40]. However, this $H_{mp}$ of Eq. (2) is hard to be linearized for a collinear magnet system, since it consists of at least one phonon and two magnon operators [26]. It is still unclear how this cubic term modulates the topological properties in magnon [31–33]; thus, the current magnon Hall theory based on the LSWT [7] would fail to give a realistic answer. So, we would like to suggest at the moment that the significant suppression of $\kappa_{xy}^{mag}$ in CGT is likely to originate from the magnon damping effect. One interesting point is that the overall temperature dependence of $\kappa_{xy}^{mag}(T)$ is still similar to $\kappa_{xy}^{calc}(T)$ obtained from LSWT.

## IV. SUMMARY

In summary, we measured the thermal Hall effect in CGT under the out-of-plane magnetic field. We observed multiple peaks in $\kappa_{xy}(T)$ with the sign change, and found that $\kappa_{xy}(H)$ follows the form of $M(H)$ with the addition of negative field-linear behavior. We demonstrated that the empirical two-component model of the following relation $\kappa_{xy}(H) = \alpha M(H) - \beta H$ fits our data exceptionally well for the overall temperature range, where $\alpha$ and $\beta$ are positive fitting parameters. We suggest that the positive magnetization-like term represents the phonon contribution $\kappa_{xy}^{ph} = \alpha M(H)$ and the negative field-linear term represents the magnon part $\kappa_{xy}^{mag} = -\beta H$. Interestingly, we noted that the temperature dependence of decomposed $\kappa_{xy}^{mag}(T)$ is similar to the theoretical magnon Hall conductivity $\kappa_{xy}^{calc}(T)$ calculated from LSWT, but the size of our $\kappa_{xy}^{mag}$ is 20 times smaller than $\kappa_{xy}^{calc}$. We interpret this significant suppression as a consequence of magnon damping due to strong spin-phonon coupling in CGT. Our results provide experimental evidence of how the topological properties of bosonic systems get affected by beyond-quadratic terms in the Hamiltonian.

## ACKNOWLEDGMENTS

We thank Pyeongjae Park, Chaebin Kim, and Jae-Ho Chung for their valuable discussions. The work at SNU was supported by the Leading Researcher Program of Korea's National Research Foundation (Grant No. 2020R1A3B2079375).

**Reference**

[1] M. V. Berry, Proc. R. Soc. London. A. Math. Phys. Sci. **392**, 45 (1984).

[2] D. J. Thouless, M. Kohmoto, M. P. Nightingale, and M. den Nijs, Phys. Rev. Lett. **49**, 405 (1982).

[3] M. Kohmoto, Ann. Phys. (N. Y). **160**, 343 (1985).

[4] N. Nagaosa, J. Sinova, S. Onoda, A. H. MacDonald, and N. P. Ong, Rev. Mod. Phys. **82**, 1539 (2010).

[5] D. Xiao, M.-C. Chang, and Q. Niu, Rev. Mod. Phys. **82**, 1959 (2010).

[6] H. Katsura, N. Nagaosa, and P. A. Lee, Phys. Rev. Lett. **104**, 066403 (2010).

[7] R. Matsumoto and S. Murakami, Phys. Rev. Lett. **106**, 197202 (2011).

[8] I. Dzyaloshinsky, J. Phys. Chem. Solids **4**, 241 (1958).

[9] T. Moriya, Phys. Rev. **120**, 91 (1960).

[10] A. Mook, J. Henk, and I. Mertig, Phys. Rev. B **89**, 134409 (2014).

[11] Y. Onose, T. Ideue, H. Katsura, Y. Shiomi, N. Nagaosa, and Y. Tokura, Science (80-. ). **329**, 297 (2010).

[12] T. Ideue, Y. Onose, H. Katsura, Y. Shiomi, S. Ishiwata, N. Nagaosa, and Y. Tokura, Phys. Rev. B **85**, 134411 (2012).

[13] M. Hirschberger, R. Chisnell, Y. S. Lee, and N. P. Ong, Phys. Rev. Lett. **115**, 106603 (2015).

[14] H. Lee, J. H. Han, and P. A. Lee, Phys. Rev. B **91**, 125413 (2015).

[15] D. Watanabe, K. Sugii, M. Shimozawa, Y. Suzuki, T. Yajima, H. Ishikawa, Z. Hiroi, T. Shibauchi, Y. Matsuda, and M. Yamashita, Proc. Natl. Acad. Sci. **113**, 8653 (2016).

[16] H. Doki, M. Akazawa, H.-Y. Lee, J. H. Han, K. Sugii, M. Shimozawa, N. Kawashima, M. Oda, H. Yoshida, and M. Yamashita, Phys. Rev. Lett. **121**, 097203 (2018).

[17] M. Yamashita, M. Akazawa, M. Shimozawa, T. Shibauchi, Y. Matsuda, H. Ishikawa, T. Yajima, Z. Hiroi, M. Oda, H. Yoshida, H.-Y. Lee, J. H. Han, and N. Kawashima, J. Phys. Condens. Matter **32**, 074001 (2020).

[18] M. Akazawa, M. Shimozawa, S. Kittaka, T. Sakakibara, R. Okuma, Z. Hiroi, H.-Y. Lee, N. Kawashima, J. H. Han, and M. Yamashita, Phys. Rev. X **10**, 041059 (2020).

[19] J. Fransson, A. M. Black-Schaffer, and A. V. Balatsky, Phys. Rev. B **94**, 075401 (2016).

[20] S. A. Owerre, J. Phys. Condens. Matter **28**, 386001 (2016).

[21] S. K. Kim, H. Ochoa, R. Zarzuela, and Y. Tserkovnyak, Phys. Rev. Lett. **117**, 227201 (2016).

[22] F. D. M. Haldane, Phys. Rev. Lett. **61**, 2015 (1988).

[23] H. Zhang, C. Xu, C. Carnahan, M. Sretenovic, N. Suri, D. Xiao, and X. Ke, Phys. Rev. Lett. **127**, 247202 (2021).

[24] J. Oh, M. D. Le, J. Jeong, J. Lee, H. Woo, W.-Y. Song, T. G. Perring, W. J. L. Buyers, S.-W. Cheong, and J.-G. Park, Phys. Rev. Lett. **111**, 257202 (2013).

[25] K. Park, J. Oh, J. C. Leiner, J. Jeong, K. C. Rule, M. D. Le, and J.-G. Park, Phys. Rev. B **94**, 104421 (2016).

[26]   T. Kim, K. Park, J. C. Leiner, and J.-G. Park, J. Phys. Soc. Japan **88**, 081003 (2019).

[27]   A. L. Chernyshev and M. E. Zhitomirsky, Phys. Rev. Lett. **97**, 207202 (2006).

[28]   A. L. Chernyshev and M. E. Zhitomirsky, Phys. Rev. B **79**, 144416 (2009).

[29]   J. Oh, M. D. Le, H.-H. Nahm, H. Sim, J. Jeong, T. G. Perring, H. Woo, K. Nakajima, S. Ohira-Kawamura, Z. Yamani, Y. Yoshida, H. Eisaki, S. W. Cheong, A. L. Chernyshev, and J.-G. Park, Nat. Commun. **7**, 13146 (2016).

[30]   K. Park, J. Oh, K. H. Lee, J. C. Leiner, H. Sim, H.-H. Nahm, T. Kim, J. Jeong, D. Ishikawa, A. Q. R. Baron, and J.-G. Park, Phys. Rev. B **102**, 085110 (2020).

[31]   A. L. Chernyshev and P. A. Maksimov, Phys. Rev. Lett. **117**, 187203 (2016).

[32]   S. Murakami and A. Okamoto, J. Phys. Soc. Japan **86**, 011010 (2017).

[33]   P. A. McClarty, Annu. Rev. Condens. Matter Phys. **13**, 171 (2022).

[34]   Y. Liu and C. Petrovic, Phys. Rev. B **96**, 054406 (2017).

[35]   Y. Tian, M. J. Gray, H. Ji, R. J. Cava, and K. S. Burch, 2D Mater. **3**, 025035 (2016).

[36]   S. Spachmann, A. Elghandour, S. Selter, B. Büchner, S. Aswartham, and R. Klingeler, Phys. Rev. Res. **4**, L022040 (2022).

[37]   S. Spachmann, S. Selter, B. Büchner, S. Aswartham, and R. Klingeler, arXiv:2207.04020 (2022).

[38]   Y. Liu, M. Han, Y. Lee, M. O. Ogunbunmi, Q. Du, C. Nelson, Z. Hu, E. Stavitski, D. Graf, K. Attenkofer, S. Bobev, L. Ke, Y. Zhu, and C. Petrovic, Adv. Funct. Mater. **32**, 2105111 (2022).

[39]   F. Zhu, L. Zhang, X. Wang, F. J. dos Santos, J. Song, T. Mueller, K. Schmalzl, W. F. Schmidt, A. Ivanov, J. T. Park, J. Xu, J. Ma, S. Lounis, S. Blügel, Y. Mokrousov, Y. Su, and T. Brückel, Sci. Adv. **7**, 1 (2021).

[40]   L. Chen, C. Mao, J.-H. Chung, M. B. Stone, A. I. Kolesnikov, X. Wang, N. Murai, B. Gao, O. Delaire, and P. Dai, Nat. Commun. **13**, 4037 (2022).

[41]   $S_{i,l}$ is i-th spin on l-th layer and $J_n(J_{c,n})$ is intra(inter)-layer exchange parameters of each n-th NN bonds. $\langle ij \rangle_n$ indicates the set of n-th NN bonds and $\langle lm \rangle$ indicates the set of the NN layers. $D$ is the DM vector pointing out of the honeycomb plane, $K$ is the easy-axis anisotropy parameter, $\mu_B$ is Bohr magneton, and $g$ is g-factor. Here we set g-factor as 2. In this paper, we defined the positive direction of $D$ as pointing out from the honeycomb plane with clockwise arranged second NN bonds (see Fig. 1(a)).

[42]   J. Zeisner, A. Alfonsov, S. Selter, S. Aswartham, M. P. Ghimire, M. Richter, J. van den Brink, B. Büchner, and V. Kataev, Phys. Rev. B **99**, 165109 (2019).

[43]   H.-L. Kim, M. J. Coak, J. C. Baglo, K. Murphy, R. W. Hill, M. Sutherland, M. C. Hatnean, G. Balakrishnan, and J.-G. Park, Rev. Sci. Instrum. **90**, 103904 (2019).

[44]   C. Tinsman, G. Li, C. Su, T. Asaba, B. Lawson, F. Yu, and L. Li, Appl. Phys. Lett. **108**, 261905 (2016).

[45]   K. Sugii, M. Shimozawa, D. Watanabe, Y. Suzuki, M. Halim, M. Kimata, Y. Matsumoto, S. Nakatsuji, and M. Yamashita, Phys. Rev. Lett. **118**, 145902 (2017).

[46]   R. Hentrich, M. Roslova, A. Isaeva, T. Doert, W. Brenig, B. Büchner, and C. Hess, Phys. Rev. B **99**, 085136 (2019).

[47]   S. Sim, H. Yang, H.-L. Kim, M. J. Coak, M. Itoh, Y. Noda, and J.-G. Park, Phys. Rev. Lett. **126**, 015901 (2021).

[48] A. Haglund, Ph.D. Thesis, Univ. Tennessee, Knoxv. (2019).

[49] R. Berman, *Thermal Conductions in Solids* (Clarendon Press, Oxford, 1976).

[50] P. A. Sharma, J. S. Ahn, N. Hur, S. Park, S. B. Kim, S. Lee, J.-G. Park, S. Guha, and S.-W. Cheong, Phys. Rev. Lett. **93**, 177202 (2004).

[51] L. D. Casto, A. J. Clune, M. O. Yokosuk, J. L. Musfeldt, T. J. Williams, H. L. Zhuang, M.-W. Lin, K. Xiao, R. G. Hennig, B. C. Sales, J.-Q. Yan, and D. Mandrus, APL Mater. **3**, 041515 (2015).

[52] D. Bansal, J. L. Niedziela, A. F. May, A. Said, G. Ehlers, D. L. Abernathy, A. Huq, M. Kirkham, H. Zhou, and O. Delaire, Phys. Rev. B **95**, 054306 (2017).

[53] C. A. Pocs, I. A. Leahy, H. Zheng, G. Cao, E.-S. Choi, S.-H. Do, K.-Y. Choi, B. Normand, and M. Lee, Phys. Rev. Res. **2**, 013059 (2020).

[54] J. Callaway, Phys. Rev. **113**, 1046 (1959).

[55] Y. Liu, R. A. Susilo, Y. Lee, A. M. M. Abeykoon, X. Tong, Z. Hu, E. Stavitski, K. Attenkofer, L. Ke, B. Chen, and C. Petrovic, ACS Nano **16**, 13134 (2022).

[56] A. Noah, H. Alpern, S. Singh, A. Gutfreund, G. Zisman, T. D. Feld, A. Vakahi, S. Remennik, Y. Paltiel, M. E. Huber, V. Barrena, H. Suderow, H. Steinberg, O. Millo, and Y. Anahory, Nano Lett. **22**, 3165 (2022).

[57] C. Herring, Phys. Rev. **95**, 954 (1954).

[58] M. A. Weilert, M. E. Msall, A. C. Anderson, and J. P. Wolfe, Phys. Rev. Lett. **71**, 735 (1993).

[59] D. Bugallo, E. Langenberg, E. Ferreiro-Vila, E. H. Smith, C. Stefani, X. Batlle, G. Catalan, N. Domingo, D. G. Schlom, and F. Rivadulla, ACS Appl. Mater. Interfaces **13**, 45679 (2021).

[60] See Supplemental materials, which include Ref. [80,81], for the reliability of thermal Hall measurement, additional fitting results, high field $\kappa_{xy}(H)$ data, and magnon Hall conductivity calculation.

[61] X. Zhang, Y. Zhang, S. Okamoto, and D. Xiao, Phys. Rev. Lett. **123**, 167202 (2019).

[62] S. Park and B.-J. Yang, Phys. Rev. B **99**, 174435 (2019).

[63] G. Go, S. K. Kim, and K.-J. Lee, Phys. Rev. Lett. **123**, 237207 (2019).

[64] S. Zhang, G. Go, K.-J. Lee, and S. K. Kim, Phys. Rev. Lett. **124**, 147204 (2020).

[65] G. Grissonnanche, A. Legros, S. Badoux, E. Lefrançois, V. Zatko, M. Lizaire, F. Laliberté, A. Gourgout, J.-S. Zhou, S. Pyon, T. Takayama, H. Takagi, S. Ono, N. Doiron-Leyraud, and L. Taillefer, Nature **571**, 376 (2019).

[66] G. Grissonnanche, S. Thériault, A. Gourgout, M.-E. Boulanger, E. Lefrançois, A. Ataei, F. Laliberté, M. Dion, J.-S. Zhou, S. Pyon, T. Takayama, H. Takagi, N. Doiron-Leyraud, and L. Taillefer, Nat. Phys. **16**, 1108 (2020).

[67] M.-E. Boulanger, G. Grissonnanche, S. Badoux, A. Allaire, É. Lefrançois, A. Legros, A. Gourgout, M. Dion, C. H. Wang, X. H. Chen, R. Liang, W. N. Hardy, D. A. Bonn, and L. Taillefer, Nat. Commun. **11**, 5325 (2020).

[68] L. Chen, M.-E. Boulanger, Z.-C. Wang, F. Tafti, and L. Taillefer, Proc. Natl. Acad. Sci. **119**, 1 (2022).

[69] L. Zhang, J. Ren, J.-S. Wang, and B. Li, Phys. Rev. Lett. **105**, 225901 (2010).

[70] T. Qin, J. Zhou, and J. Shi, Phys. Rev. B **86**, 104305 (2012).

[71] T. Saito, K. Misaki, H. Ishizuka, and N. Nagaosa, Phys. Rev. Lett. **123**, 255901 (2019).

[72] M. Mori, A. Spencer-Smith, O. P. Sushkov, and S. Maekawa, Phys. Rev. Lett. **113**, 265901 (2014).

[73] J.-Y. Chen, S. A. Kivelson, and X.-Q. Sun, Phys. Rev. Lett. **124**, 167601 (2020).

[74] L. Mangeolle, L. Balents, and L. Savary, arXiv:2202.10366 (2022).

[75] H. Guo, D. G. Joshi, and S. Sachdev, arXiv:2201.11681 (2022).

[76] Y. Kagan and L. A. Maksimov, Phys. Rev. Lett. **100**, 145902 (2008).

[77] C. Strohm, G. L. J. A. Rikken, and P. Wyder, Phys. Rev. Lett. **95**, 155901 (2005).

[78] Y. Yang, G.-M. Zhang, and F.-C. Zhang, Phys. Rev. Lett. **124**, 186602 (2020).

[79] S.-H. Do, J. A. M. Paddison, G. Sala, T. J. Williams, K. Kaneko, K. Kuwahara, A. F. May, J. Yan, M. A. McGuire, M. B. Stone, M. D. Lumsden, and A. D. Christianson, Phys. Rev. B **106**, L060408 (2022).

[80] P. Czajka, T. Gao, M. Hirschberger, P. Lampen-Kelley, A. Banerjee, J. Yan, D. G. Mandrus, S. E. Nagler, and N. P. Ong, Nat. Phys. **17**, 915 (2021).

[81] T. Fukui, Y. Hatsugai, and H. Suzuki, J. Phys. Soc. Japan **74**, 1674 (2005).

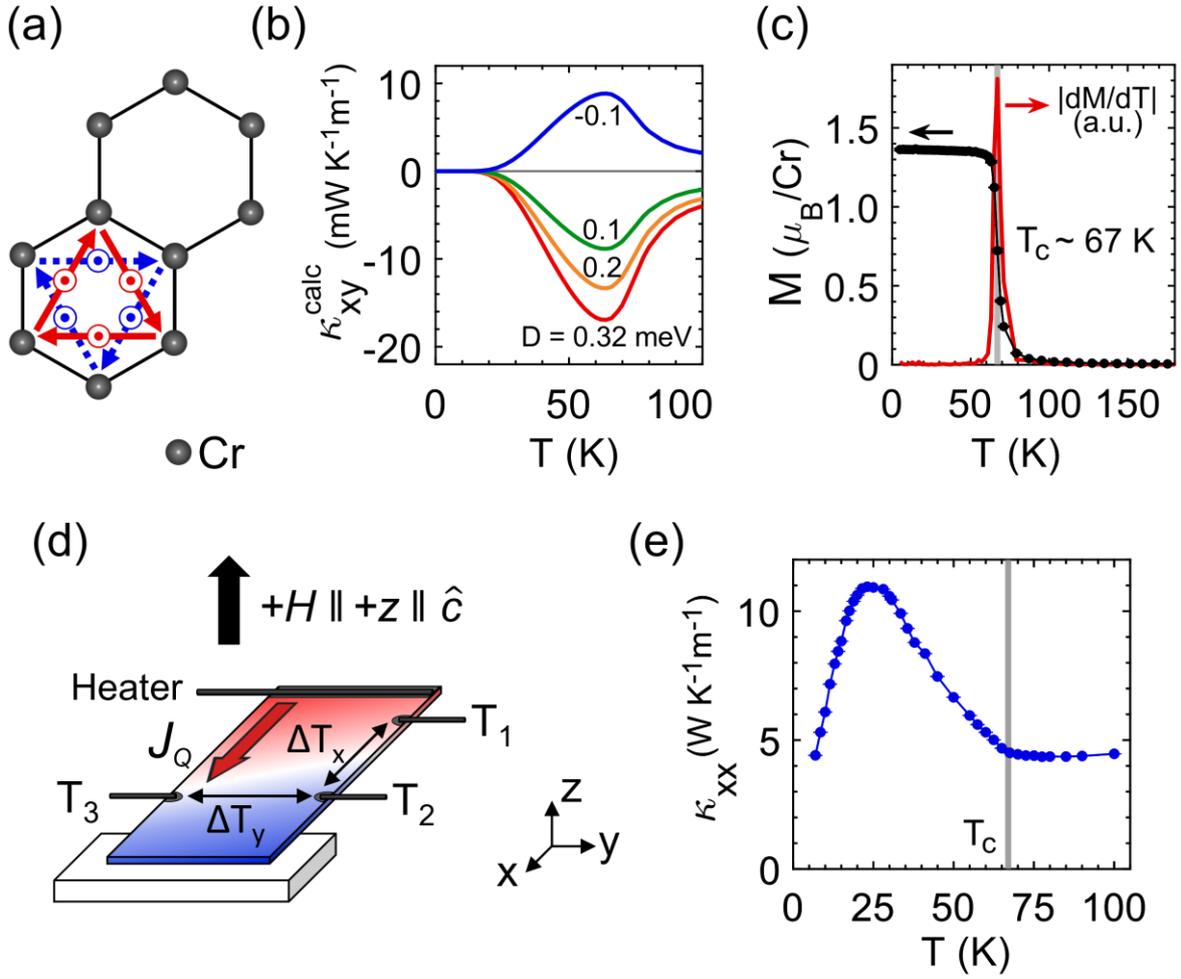

**FIG. 1.** (a) Cr - based honeycomb structure in $Cr_2Ge_2Te_6$. Each arrow indicates second-NN bonds for DM interaction (***D***). The positive direction of ***D*** is pointing out from the honeycomb plane with clockwise arranged second NN-bonds. (b) Theoretical magnon Hall conductivity $\kappa_{xy}^{calc}(T)$ with different $|D|$ values, calculated from the parameter set A in Table I. (c) The temperature dependence of out-of-plane magnetization ($M$) and $|dM/dT|$ with magnetic field of 0.1 T. (d) Schematic of thermal Hall measurement setup. $J_Q$ denotes heat current. (e) Temperature dependence of $\kappa_{xx}$ under zero field.

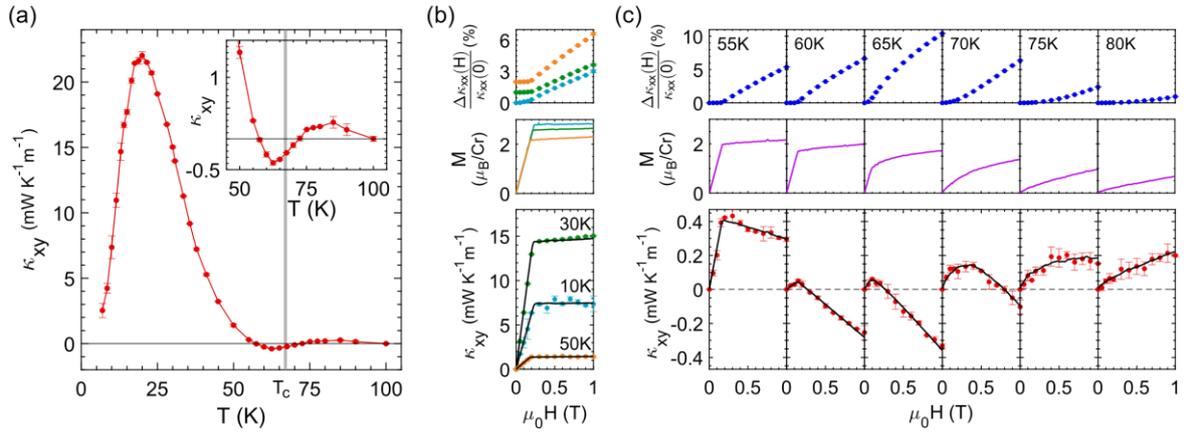

**FIG. 2.** (a) Temperature dependence of $\kappa_{xy}$ under the field of 1 T. The inset shows the blown-up picture of the sign change in $\kappa_{xy}$ near $T_C$. (b) $\Delta\kappa_{xx}(H)/\kappa_{xx}(0)$, $M(H)$, and $\kappa_{xy}(H)$ at $T$ = 10, 30, and 50 K. $\Delta\kappa_{xx}(H)/\kappa_{xx}(0)$ data are shifted upward for better presentation. (c) $\Delta\kappa_{xx}(H)/\kappa_{xx}(0)$, $M(H)$, and $\kappa_{xy}(H)$ at $T$ = 55, 60, 65, 70, 75, and 80 K. Solid black curves are fitting results obtained from the empirical two-component model $\kappa_{xy}(H) = \alpha M(H) - \beta H, (\alpha, \beta \geq 0)$.

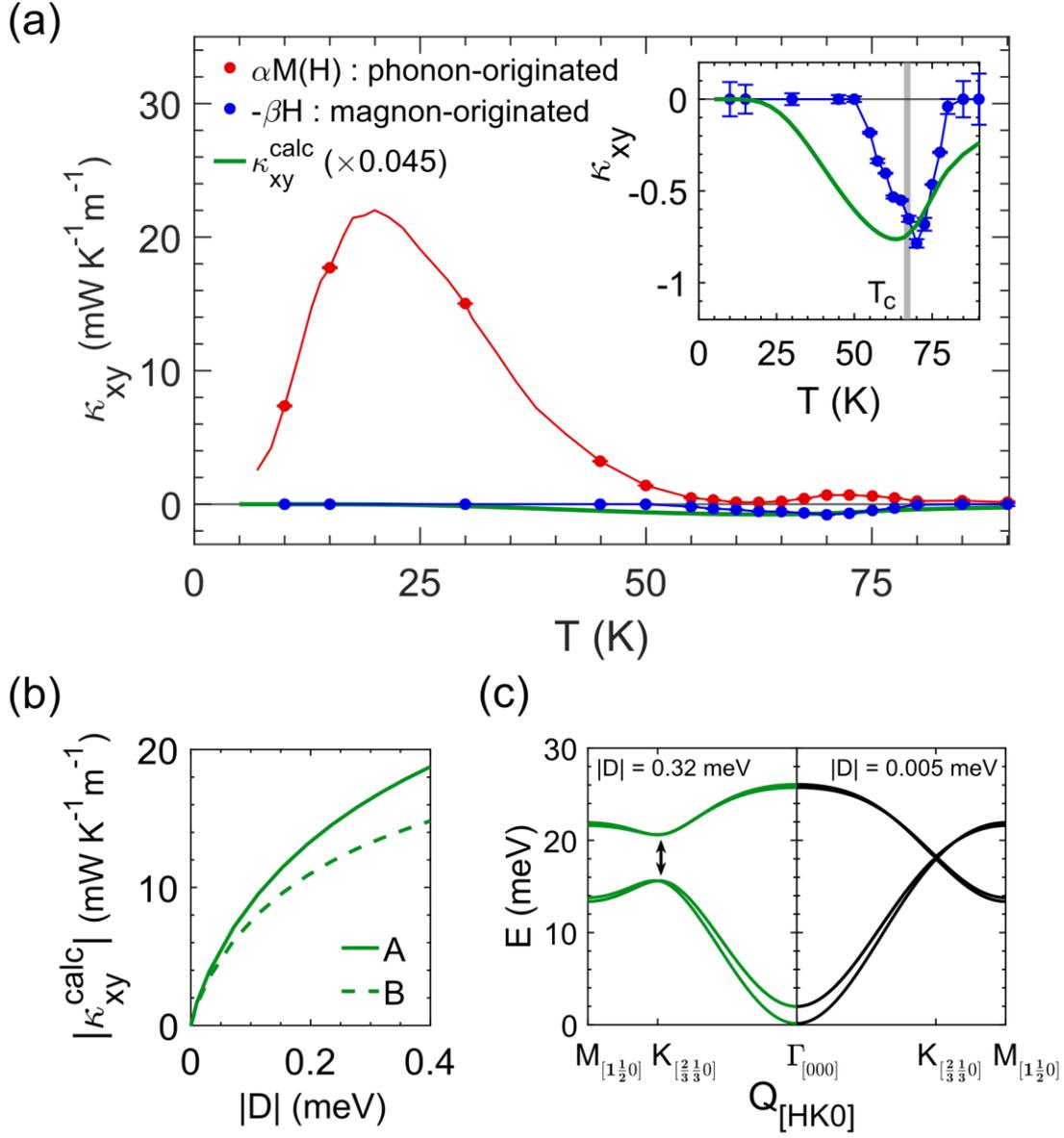

**FIG. 3.** (a) Temperature dependence of $\alpha M(H)$ term (red dots, phonon-originated term), and $-\beta H$ term (blue dots, magnon-originated term) under the field of 1 T. Red and blue curves are guide to the eyes. A solid green curve is a theoretical $\kappa_{xy}^{calc}(T)$ calculated from the parameter set A in Table I, which is multiplied by a factor of 0.045. Inset is the blown-up picture showing the overall temperature dependence of the magnon-originated term ($-\beta H$) and the scaled $\kappa_{xy}^{calc}(T)$. (b) $|\kappa_{xy}^{calc}|$ as a function of the size of DM interaction at $T$ = 64 K, calculated from both parameter sets A and B in Table I. (c) Magnon dispersion calculated from parameter set A in Table I with two different DM interactions, $|D|$ = 0.005, and 0.32 meV.

| Label | A | B |
|---|---|---|
| $J_1$ | -2.73 | -2.76 |
| $J_2$ | -0.33 | -0.11 |
| $J_3$ | 0 | -0.33 |
| $J_{c,1}$ | -0.10 | -0.86 |
| $J_{c,2}$ | -0.08 | 0 |
| $|D|$ | 0.32 | 0.20 |
| $K$ | 0.01 | 0.033 |
| Reference | [39] | [40] |

TABLE I. Two parameter sets of magnetic Hamiltonian ($H_m$) for CGT obtained from recent neutron studies [39,40]. All the parameters are given in the unit of meV.